\documentclass[final, journal, letterpaper, oneside, twocolumn]{IEEEtran}
\usepackage{cite}
\usepackage{graphicx}
\usepackage{amssymb,amsmath}
\usepackage{amsfonts}
\usepackage{multirow}
\usepackage{mathrsfs}
\usepackage{color}
\usepackage{url}
\usepackage{flushend}
\usepackage{setspace}
\usepackage{placeins}
\allowdisplaybreaks

\newtheorem{lemma}{Lemma}
\newtheorem{theorem}{Theorem}

\begin{document}

\title{Spectrum-Sharing Multi-Hop Cooperative Relaying: Performance Analysis Using Extreme Value Theory}

\author{Minghua~Xia,~\IEEEmembership{Member,~IEEE}, and Sonia~A\"{\i}ssa,~\IEEEmembership{Senior Member,~IEEE}
\thanks{Manuscript received 25 February, 2013; revised 16 June and 11 August, 2013; accepted 11 September, 2013. The associate editor coordinating the review of this paper and approving it for publication was D. Niyato.}
\thanks{This work was supported by a Discovery Accelerator Supplement (DAS) Grant from the Natural Sciences and Engineering Research Council (NSERC) of Canada.}
\thanks{The authors are with the Institut National de la Recherche Scientifique--Energy, Materials and Telecommunications center (INRS-EMT), University of Quebec, Montreal, QC, H5A 1K6, Canada (e-mail: \{xia, aissa\}@emt.inrs.ca).}
\thanks{Digital Object Identifier XXX}
}

\markboth{IEEE TRANSACTIONS ON WIRELESS COMMUNICATIONS, ACCEPTED FOR PUBLICATION} {XIA \MakeLowercase{\textit{et al.}}: Spectrum-Sharing Multi-Hop Cooperative Relaying: Performance Analysis Using Extreme Value Theory}

\maketitle

\IEEEpubid{XXX~\copyright~2013 IEEE}

\begin{abstract}
\noindent In spectrum-sharing cognitive radio systems, the transmit power of secondary users has to be very low due to the restrictions on the tolerable interference power dictated by primary users. In order to extend the coverage area of secondary transmission and reduce the corresponding interference region, multi-hop amplify-and-forward (AF) relaying can be implemented for the communication between secondary transmitters and receivers. This paper addresses the fundamental limits of this promising technique. Specifically, the effect of major system parameters on the performance of spectrum-sharing multi-hop AF relaying is investigated. To this end, the optimal transmit power allocation at each node along the multi-hop link is firstly addressed. Then, the extreme value theory is exploited to study the limiting distribution functions of the lower and upper bounds on the end-to-end signal-to-noise ratio of the relaying path. Our results disclose that the diversity gain of the multi-hop link is always unity, regardless of the number of relaying hops. On the other hand, the coding gain is proportional to the water level of the optimal water-filling power allocation at secondary transmitter and to the large-scale path-loss ratio of the desired link to the interference link at each hop, yet is inversely proportional to the accumulated noise, i.e. the product of the number of relays and the noise variance, at the destination. These important findings do not only shed light on the performance of the secondary transmissions but also benefit system designers improving the efficiency  of future spectrum-sharing cooperative systems.
\end{abstract}


\IEEEpubidadjcol

\section{Introduction}
\label{Section:Introduction}
Cognitive radio (CR) is a promising wireless technology to resolve the growing scarcity of the indispensable electromagnetic spectrum resources. By use of CR, secondary users (SUs) without explicitly assigned spectrum resources can co-exist with primary users (PUs) licensed with particular spectrum. In practice, some major communications regulators like the Federal Communications Committee (FCC) in U.S. and the Office of Communications (OFCOM) in U.K. have allowed secondary access for unlicensed devices to the terrestrial TV broadcast bands. Among various forms of CR implementation, spectrum-sharing CR is especially appealing for practical deployment since it does not involve complex spectrum-sensing mechanisms. More specifically, spectrum-sharing CR limits only the transmit power of SUs such that their harmful interference onto PUs remains below prescribed tolerable levels \cite{FCC2002}.

Because of the interference power constraint dictated by PUs, the transmit power of SUs in spectrum-sharing systems has to be very low, which limits the coverage area of secondary transmission. In order to extend the coverage area of secondary transmission and guarantee reliable communication, cooperative relaying techniques can be exploited. Using relaying techniques, a single or multiple idle users (either PUs or SUs) can be involved in forwarding messages between a secondary source and its destination. The relaying can be implemented using two main techniques, amplify-and-forward (AF) and decode-and-forward (DF). In particular, assuming DF protocol and simple dual-hop transmission, system performance of dual-hop relaying in spectrum-sharing systems was extensively studied, see e.g. \cite{AsghariICC10, MusavianTWC10May, LeeTWC11Feb} and references therein. Nevertheless, apart from its implementation simplicity, the AF relaying can outperform the DF protocol when a relay is closer to its receiver \cite[p.~11]{E1Gamal2011}. Until lately, dual-hop AF relaying was integrated into spectrum-sharing systems and its performance was explored \cite{XiaTCOM12June, XiaTCOM12Nov}.

\IEEEpubidadjcol

In order to further extend the coverage area of the secondary transmitters yet reduce their interference region and its negative impact on PUs, multi-hop AF relaying can be implemented in real-world spectrum-sharing systems. For an efficient implementation of such techniques, performance evaluation and design are necessary. However, when analyzing system performance of multi-hop relaying, the simplistic dual-hop model does not capture many important features of multi-hop systems. For example, from the viewpoint of SUs, system performance of multi-hop relaying will be inevitably affected by the interference-power constraint dictated by PUs, which should be guaranteed by all intermediate nodes between a secondary source and its final destination. In order to mitigate this constraint, dynamic power allocation at the secondary transmitters is necessary, but how about the effect of power allocation on the end-to-end performance of secondary multi-hop link? In particular, compared to the simple dual-hop case, how will the increasing number of relaying hops affect system performance? These issues are critical for an efficient implementation and deployment of multi-hop relaying in spectrum-sharing systems and will be explicitly addressed in this paper.

Accurate performance analysis of spectrum-sharing multi-hop AF relaying is hardly tractable in closed-form due to extremely high mathematical difficulty. Actually, even for the conventional multi-hop AF relaying without spectrum sharing, closed-form performance analysis is still an open problem. In this work, we investigate the lower and upper bounds on the end-to-end signal-to-noise ratio (SNR) of spectrum-sharing multi-hop AF relaying systems. In order to gain illuminating insights into system performance, we exploit extreme value theory and study limiting distribution functions of the said bounds, which are further applied to analyze the performance in terms of the outage probability, the diversity gain, the coding gain, and the achievable data rate. In particular, our results disclose that the diversity gain of the system under study is always unity, irrespective of the number of relaying hops. On the other hand, the coding gain is proportional to the water level of the optimal water-filling power allocation at each secondary transmitter along the multi-hop link and to the large-scale path-loss ratio of the desired link to the interference link at each hop, yet is inversely proportional to the accumulated noise at the destination, namely, the product of the number of relays and noise variance.

In detailing the above highlighted contributions, the rest of this paper is organized as follows. Section~\ref{Section:SystemModel} describes the system model. Section~\ref{Section:PowerAllocation} is devoted to the optimal transmit power allocation. Subsequently, in Section~\ref{Section:e2eSNR}, the exact distribution functions of the end-to-end SNR are derived and, for further processing, the limiting distribution functions of the lower and upper bounds on the end-to-end SNR follow. With the resultant distribution functions, the system performance is extensively analyzed and discussed in Section~\ref{Section:PerformanceAnalysis}. Concluding remarks are provided in Section~\ref{Section:Conclusion} and, finally, some mathematical tools and detailed derivations are relegated to appendices.

\section{System Model and Assumptions}
\label{Section:SystemModel}
\begin{figure}[t]
\centering
\includegraphics [width=2.2in, clip, keepaspectratio]{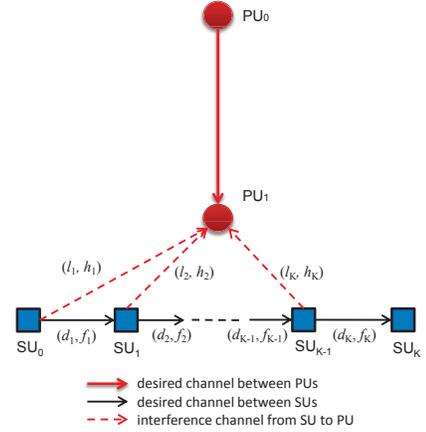}
\caption{System model of spectrum-sharing based secondary multi-hop relaying, where the primary transmitter $\mathrm{PU}_0$ is located far from the secondary users $\mathrm{SU}_k, k \in [0, K]$, compared to the primary receiver $\mathrm{PU}_1$. Parameters $(d_k, f_k)$ denote the distance and the channel fast-fading coefficient between $\mathrm{SU}_{k-1}$ and $\mathrm{SU}_{k}$ (desired link). Similarly, $(l_k, h_k)$ refer to the distance and the channel coefficient between $\mathrm{SU}_{k-1}$ and $\mathrm{PU}_{1}$ (interference link).}
\label{Fig.SystemModel}
\end{figure}

We consider a $K$-hop cooperative relaying system operating in a spectrum-sharing CR environment. As depicted in Fig.~\ref{Fig.SystemModel}, secondary users $\mathrm{SU}_0$ and $\mathrm{SU}_{K}$ exchange data with the assistance of $K-1$ SUs serving as consecutive AF relays, while primary source $\mathrm{PU}_0$ is transmitting to primary receiver $\mathrm{PU}_1$. It is assumed that all nodes are equipped with a single half-duplex omnidirectional antenna each. In order to share the spectrum resources originally licensed to PUs, the transmit power of SUs is strictly limited by the prescribed tolerable interference power at the primary receiver. More details on the optimal power allocation at the secondary nodes along the multi-hop relaying link will be given in the next section.

For the secondary multi-hop AF relaying link, all SUs work in a time-division multiple access (TDMA) fashion and equal time slots of a transmission frame are allocated to each SU along the multi-hop path. Also, only one SU transmits to its next node along the multi-hop relaying path during each time slot. With these assumptions in mind, based on \cite{HasnaCL03May, KaragiannidisTCOM06Jan}, the end-to-end (e2e) received SNR at the final destination $\mathrm{SU}_K$, can be expressed as
\begin{equation} \label{Eq.e2eSNR}
\gamma_{e2e} = \left(\sum_{k=1}^{K}\frac{1}{\gamma_k}\right)^{-1},
\end{equation}
where $\gamma_k$ denotes the received SNR at the $k^\mathrm{\mathrm{th}}$ secondary node ($k=1, \cdots, K$), defined as
\begin{equation} \label{Eq.SNR@Relay}
\gamma_k \triangleq \frac{P_{k-1}}{\sigma_k^2}\,d_k^{-\epsilon}|f_k|^2,
\end{equation}
where $d_k$ and $f_k$ denote the distance and the channel fast-fading coefficient between consecutive secondary nodes $\mathrm{SU}_{k-1}$ and $\mathrm{SU}_{k}$ (cf. Fig.~\ref{Fig.SystemModel}), $\epsilon \ge 2$ refers to the path-loss exponent (generally, $\epsilon = 2$ in free space and $\epsilon = 4$ in most practical scenarios), $P_{k-1}$ indicates the transmit power at $\mathrm{SU}_{k-1}$, and $\sigma_k^2$ stands for the additive white Gaussian noise (AWGN) variance at $\mathrm{SU}_{k}$. For notational convenience, the secondary source node in Fig.~\ref{Fig.SystemModel} is denoted by $\mathrm{SU}_0$.

Compared to conventional multi-hop relaying transmission without spectrum sharing where the transmit power of the source or any relaying node is generally fixed, in a spectrum-sharing context, SUs share the spectrum resources originally licensed to PUs and, thus, the transmit power $P_{k}$ of the $k^\mathrm{\mathrm{th}}$ secondary node ($\mathrm{SU}_k$), $k=0,\cdots,K-1$, is strictly limited by the prescribed tolerable interference power at the primary receiver. Moreover, it is seen from \eqref{Eq.e2eSNR}-\eqref{Eq.SNR@Relay} that the actual power values will greatly affect the end-to-end SNR at the destination node $\mathrm{SU}_K$. Therefore, in the next section we establish the optimal transmit power at $\mathrm{SU}_k$.

\textbf{Remark 1 (Relay Gain):}
For a multi-hop relaying system, if the relay gain is set to the inverse of the channel gain of the previous hop, Eq. \eqref{Eq.e2eSNR} is the exact expression of the end-to-end SNR (cf. the paragraphs after \cite[Eq. (4)]{HasnaCL03May} and \cite[Eq. (11)]{KaragiannidisTCOM06Jan}). On the other hand, if the relay gain is set to the inverse of the channel gain of the previous hop plus noise variance, Eq. \eqref{Eq.e2eSNR} acts an upper bound on the end-to-end SNR. The former enables mathematical tractability and serves as a benchmark for all practical multi-hop AF relaying schemes. In particular, a comparison of the outage probability between multi-hop systems with these two different relay gains demonstrated only a slight performance difference even in the low and medium SNR regions (cf. \cite[Fig. 1]{HasnaCL03May}).

\textbf{Remark 2 (The Interference Coming From The Primary Transmitter):}
For ease of mathematical tractability, the interference coming from the primary transmitter is not separately counted but treated as noise, as shown in Eq. \eqref{Eq.SNR@Relay}. This is feasible if the interference is weak, for example, when the primary transmitter is located far from the secondary users as illustrated in Fig. \ref{Fig.SystemModel} \cite{Kang08VTCspring}. Furthermore, the interference can be treated as noise if the primary transmitter's signal is generated by random Gaussian codebooks \cite{Etkin07JSAC03}. Theoretically, the validity of the methodology treating non-Gaussian interference as noise was originally proven in \cite{Lapidoth96TIT09}. For an elaborate investigation into the effect of the interference on the end-to-end performance of dual-hop relaying, we refer the interested reader to \cite{XiaTCOM12Nov}.

\section{Optimal Power Allocation}
\label{Section:PowerAllocation}
In this section, the criterion for power allocation at the secondary nodes along the multi-hop relaying link (including the source $\mathrm{SU}_0$ and its consecutive relaying nodes $\mathrm{SU}_k$, $k=1, 2, \cdots, K-1$) is firstly established. Then, based on this criterion, the value of the optimal transmit power during each transmission slot at each secondary node is determined.

\subsection{Criterion for Power Allocation at Secondary Transmitters}
\label{Subsection:Criteria4PowerAllocation-S1}
In this part, the criterion for optimal power allocation at the secondary nodes $\mathrm{SU}_{k-1}$, $k=1, 2, \cdots, K$, is established. Assume that the average tolerable interference power at the primary receiver is $W$ dB with respect to the noise power. Also, the distances corresponding to the desired link and the interference link at each hop are assumed to remain fixed during a communication between the secondary source and its destination  (cf. Fig.~\ref{Fig.SystemModel}). Thus, to maximize the achievable data rate at the $k^\mathrm{\mathrm{th}}$ hop, we need to optimize the transmit power at $\mathrm{SU}_{k-1}$ with respect to $W$ and the instantaneous fast-fading fluctuations of the corresponding desired channel and interference channel. To this end, it is easy to show that the first-order derivative of the end-to-end received SNR $\gamma_{e2e}$ given by \eqref{Eq.e2eSNR} with respect to $\gamma_k$ is strictly positive and, hence, $\gamma_{e2e}$ is a monotonically increasing function of $\gamma_k$. Consequently, in order to maximize the achievable data rate $R_k$ at the $k^\mathrm{\mathrm{th}}$ hop, the optimal transmit power $P_{k-1}$ at $\mathrm{SU}_{k-1}$ is determined as the solution to the optimization problem:
\begin{equation}  \label{Eq.S1CapacityA}
R_k = \max\limits_{{P_{k-1}} \ge 0}\mathcal{E}_{{f_{k}}}\left\{\log_2\left(1+\frac{{P_{k-1}}}{\sigma_{k}^2}\,d_k^{-\epsilon}|f_{k}|^2\right)\right\}
\end{equation}
\begin{equation}  \label{Eq.S1CapacityB}
\mathrm{s.t.} \quad \mathcal{E}_{h_{k}}\left\{{P_{k-1}}\,l_k^{-\epsilon}|h_{k}|^2\right\} \le 10^{W/10},
\end{equation}
where $l_k$ and $h_{k}$ denote the distance and the channel fast-fading coefficient of the interference link from $\mathrm{SU}_{k-1}$ to $\mathrm{PU}_1$, and the operator $\mathcal{E}\{\,.\,\}$ means statistical expectation. Clearly, the optimal transmit power $P_{k-1}$ at secondary node $\mathrm{SU}_{k-1}$ is a function of the parameters of the desired link and the interference link (i.e. $(d_k, f_{k})$ and $(l_k, h_{k})$, respectively, as shown in Fig.~\ref{Fig.SystemModel}), the noise variance ($\sigma_k^2$), and the average tolerable interference power at the primary receiver ($W$).

In spectrum-sharing systems, the tolerable interference power at the primary receivers can be generally defined by means of average interference power, peak interference power, or both \cite{Musavian08IET06, MusavianTWC09Jan}. Also, the maximum output-power for SUs is constrained in practice. However, the effect of the maximum output-power constraint on system performance is essentially equivalent to the peak interference-power constraint, except for a scaling factor given by the interference channel gain. Furthermore, it is already demonstrated that ``imposing a constraint on the peak interference power does not yield a significant impact on the ergodic capacity as long as the average interference power is constrained'' \cite{MusavianTWC09Jan}. As a result, only the average interference-power constraint is considered in the optimization problem above.

Applying the well-known Lagrangian multiplier method \cite[Section 5.3.3]{Tse05} to \eqref{Eq.S1CapacityA}-\eqref{Eq.S1CapacityB}, it is easy to show that the optimal transmit power at secondary node $\mathrm{SU}_{k-1}$ is given by
\begin{equation} \label{Eq.S1CapacityC}
P_{k-1} = \left[\frac{\lambda_{k-1}}{l_k^{-\epsilon}|h_{k}|^2}-\frac{\sigma_{k}^2}{d_k^{-\epsilon}|f_{k}|^2}\right]^+,
\end{equation}
where the ceiling operator $[x]^+ \triangleq \max(0,~x)$. Furthermore, the power allocation parameter $\lambda_{k-1}$ in \eqref{Eq.S1CapacityC} is determined by setting the average interference-power constraint in \eqref{Eq.S1CapacityB} to equality, such that
\begin{equation} \label{Eq.S1CapacityD0}
\mathcal{E}_{f_{k},\,h_{k}}\left\{\left[\lambda_{k-1}-\sigma_{k}^2\,\frac{l_k^{-\epsilon}|h_{k}|^2}{d_k^{-\epsilon}|f_{k}|^2}\right]^+\right\} = 10^{W/10}.
\end{equation}

Now, if $\eta_k$ is defined as the large-scale path-loss ratio of the desired link to the interference link at the $k^\mathrm{th}$ hop, i.e.
\begin{equation} \label{Eq.NormalizedDistance}
\eta_k \triangleq \frac{d_k^{-\epsilon}}{l_k^{-\epsilon}},
\end{equation}
then, \eqref{Eq.S1CapacityD0} can be reformulated as
\begin{equation} \label{Eq.S1CapacityD}
\mathcal{E}_{f_{k},\,h_{k}}\left\{\left[\lambda_{k-1}-\frac{\sigma_{k}^2}{\eta_k}\,\frac{|h_{k}|^2}{|f_{k}|^2}\right]^+\right\} = 10^{W/10}.
\end{equation}

Like the well-known water-filling power allocation algorithm \cite{GoldsmithTIT97Nov}, the power allocation parameter $\lambda_{k-1}$ associated with \eqref{Eq.S1CapacityD} corresponds to the so-called water level and it will be explicitly determined in the next subsection. Remarkably, this power allocation parameter has a significant effect on the system performance, as will be discussed further in the following Section~\ref{Subsection:DiversityCodingGains}. On the other hand, the ceiling operator $[x]^+$ in \eqref{Eq.S1CapacityC} implies that the transmit power is zero if the gain of the desired channel is smaller than a lower bound, i.e. $|f_{k}^2| \le \frac{\sigma_{k}^2}{\lambda_{k-1}\,\eta_k}|h_{k}|^2$. In such a case, no data will be transmitted. In other words, only when $|f_{k}|^2 > \frac{\sigma_{k}^2}{\lambda_{k-1}\,\eta_k}|h_{k}|^2$ can $\mathrm{SU}_{k-1}$ transmit to its next node $\mathrm{SU}_{k}$. This makes the spectrum-sharing multi-hop relaying system more energy efficient than the conventional multi-hop relaying system where the transmission between two consecutive nodes is totally regardless of the channel fluctuations in between.

\textbf{Remark 3 (Channel Estimation):}
It is noteworthy that, according to \eqref{Eq.S1CapacityC}, when performing the optimal water-filling power allocation at secondary node $\mathrm{SU}_{k-1}$, the channel parameters of the desired link $(d_k, f_k)$ and the interference link $(l_k, h_k)$ have to be efficiently estimated. Among them, the parameters $(d_k, f_k)$ between $\mathrm{SU}_{k-1}$ and $\mathrm{SU}_{k}$ can be readily obtained by using the conventional training-sequence based methodology. On the other hand, the parameters $(l_k, h_k)$ between $\mathrm{SU}_{k-1}$ and primary user $\mathrm{PU}_{1}$ can be attained at $\mathrm{SU}_{k-1}$, for example, by periodically sensing the pilot signals transmitted by $\mathrm{PU}_{1}$ under the assumption of channel reciprocity \cite{ZhaoTSP08Feb}, or by indirect feedback from a third-party such as a band manager which coordinates between the primary and secondary users \cite{PehaProc09Apr}. Although the acquisition of this channel state information (CSI) requires additional cost at SUs, it enables them to strictly comply with the prescribed interference power constraint at PUs and to maximize the achievable data rate at each hop. Also, we note that in practice, there must be errors when estimating the CSI. A further inspection of this issue is beyond the scope of the paper.

\textbf{Remark 4 (On the Optimality of Power Allocation):}
For the cognitive multi-hop relaying under study, all SUs along the multi-hop link work in a TDMA fashion and there is no central node. Thus, it is impossible to collect all CSI before performing optimal power-allocation among all SUs along the multi-hop path. Actually, for the node $\mathrm{SU}_k$ along the path, it has only the CSI pertaining to the desired link $\mathrm{SU}_k \to \mathrm{SU}_{k+1}$ and the interference link $\mathrm{SU}_k \to \mathrm{PU}_1$. Accordingly, the optimal power allocation can only be performed with respect to these CSI. In other words, the optimal power allocation can only be performed within a single hop instead of the end-to-end link. This optimization strategy is essentially to ``make best efforts'' at each relaying hop.

\subsection{Optimal Transmit Power at the $(k-1)^{\mathrm{th}}$ Node}
\label{Subsection:PowerAllocation}
According to \eqref{Eq.S1CapacityD}, in order to determine the power allocation parameter $\lambda_{k-1}$, we derive the probability density function (PDF) of $V_1 \triangleq |h_{k}|^2/|f_{k}|^2$. Since all channels in the system under study are supposed to be subject to Rayleigh fading with unit-mean, it is straightforward that $|f_{k}|^2$ and $|h_{k}|^2$ are exponentially distributed with unit-mean. Hence, the PDF of $|f_{k}|^2$ and $|h_{k}|^2$ is given by
\begin{equation}  \label{Eq.PDFRayleigh}
f_X(x) = \exp(-x), \quad X \in \left\{|f_{k}|^2,\,|h_{k}|^2\right\}.
\end{equation}

Conditioning on $|f_{k}|^2$, the PDF of $V_1$ can be given by
\begin{equation} \label{Eq.PDF-V1}
f_{V_1}(x)
= \int_0^\infty{y\exp\left[-\left(x+1\right)y\right]}\,\mathrm{d}y 
=  (x+1)^{-2}. 
\end{equation}
Then, inserting \eqref{Eq.PDF-V1} into \eqref{Eq.S1CapacityD} and performing some integrations, the power allocation parameter $\lambda_{k-1}$ can be determined by
\begin{align}
10^{\frac{W}{10}}
&  =  \int_{0}^{\frac{\eta_k\lambda_{k-1}}{\bar{\gamma}\sigma_{k}^2}}\left(\lambda_{k-1}-x\,\frac{\sigma_k^2}{\eta_k}\right)f_{V_1}(x)\,\mathrm{d}x  \nonumber \\
&  =  \frac{\lambda_{k-1}\left(\eta_k\lambda_{k-1}+\sigma_k^2\right)}{\eta_k\lambda_{k-1}+\bar{\gamma}\sigma_k^2}  
-\frac{\sigma_k^2}{\eta_k} \ln\left(1+\frac{\eta_k\lambda_{k-1}}{\bar{\gamma}\sigma_k^2}\right) \label{Eq.Lambda}
\end{align}

Although $\lambda_{k-1}$ cannot be expressed in closed-form given that \eqref{Eq.Lambda} is a transcendental equation, it can be easily computed in a numerical way because \eqref{Eq.Lambda} involves only the elementary logarithmic function. With the resulting $\lambda_{k-1}$, the optimal transmit power $P_{k-1}$ at the $(k-1)^\mathrm{th}$ secondary node can be readily determined by using \eqref{Eq.S1CapacityC}. Next, we investigate the distribution functions and moment generation function (MGF) of the end-to-end SNR at the secondary destination.

\section{Statistics of the End-to-End SNR}
\label{Section:e2eSNR}
Now, we investigate the statistics of the end-to-end SNR at the secondary destination. At first, we derive the exact distribution functions and MGF of the end-to-end SNR. For further processing while gaining insight into system performance, a pair of lower and upper bounds on the SNR are proposed and their limiting distribution functions as the number of hops $K \to \infty$ are developed, by using extreme value theory.

\subsection{Exact Distribution Functions and MGF of the End-to-End SNR}
\label{Subsection:Scenario1-SNR}
Here, we derive the exact distribution functions and MGF of the end-to-end SNR at the secondary destination. By virtue of the end-to-end SNR expression shown in \eqref{Eq.e2eSNR}, in order to determine its distribution functions, we need to firstly derive the distribution functions of its component $\gamma_k$ given by \eqref{Eq.SNR@Relay}. More specifically, substituting the obtained optimal transmit power $P_{k-1}$ shown in \eqref{Eq.S1CapacityC} into the expression \eqref{Eq.SNR@Relay}, $\gamma_k$ can be rewritten as
\begin{equation} \label{Eq.gamma2PA}
\gamma_k
= \left[\frac{\lambda_{k-1}}{\sigma_{k}^2}\frac{d_k^{-\epsilon}|f_k|^2}{l_k^{-\epsilon}|h_k|^2}-1\right]^{+}
=  \left[\frac{\lambda_{k-1} \, \eta_k}{\sigma_{k}^2}\,V_2-1\right]^{+},
\end{equation}
where $V_2 \triangleq |f_k|^2/|h_k|^2$. Since $|f_k|^2$ and $|h_k|^2$ are of the same exponential distribution (cf. Eq. \eqref{Eq.PDFRayleigh}), the PDF of $V_2$ is the same as its inverse (i.e. $V_1$ defined above). That is,
\begin{equation}  \label{Eq.PDF-V2}
f_{V_2}(x) = (x+1)^{-2}.
\end{equation}
Inserting \eqref{Eq.PDF-V2} into \eqref{Eq.gamma2PA} and performing some algebraic manipulations yields the PDF of $\gamma_k$, given by (cf. Appendix \ref{Appendix-A})
\begin{equation} \label{Eq.PDFgamma-k}
f_{\gamma_k}(\gamma) = {a_k}(\gamma+a_k)^{-2},
\end{equation}
where the constant $a_k \triangleq \lambda_{k-1}\, \eta_k/\sigma_k^2 + 1 \approx \lambda_{k-1}\, \eta_k/\sigma_k^2 $. Subsequently, performing Laplace transform over \eqref{Eq.PDFgamma-k} yields the MGF of $1/\gamma_k$, namely,
\begin{eqnarray}
M_{\frac{1}{\gamma_k}}(s)
&  =   & \int_0^\infty{\exp\left(-\frac{s}{\gamma}\right){a_k}(\gamma+a_k)^{-2}}\,\mathrm{d}\gamma  \nonumber\\
&  =   & \frac{1}{a_k} \int_0^\infty{\exp(-sx){\left(x+\frac{1}{a_k}\right)^{-2}}}\,\mathrm{d}\gamma  \label{Eq.MGFgamma-k-a}\\
&  =   & \Psi\left(1, 0; \frac{s}{a_k}\right) \label{Eq.MGFgamma-k-b},
\end{eqnarray}
where the change of variable $x=1/\gamma$ was used to reach \eqref{Eq.MGFgamma-k-a} and \cite[vol.1, Eq.(2.3.6.9)]{Prudnikov86} was employed to derive \eqref{Eq.MGFgamma-k-b} with $\Psi(a, b; x)$ being the Tricomi confluent hypergeometric function \cite[Eq.(9.210.2)]{Gradshteyn07}. Moreover, in light of \eqref{Eq.e2eSNR}, it is clear that the end-to-end SNR can be reformulated as
\begin{equation} \label{Eq.e2eSNR-2}
\frac{1}{\gamma_{e2e}} = \sum_{k=1}^{K}\frac{1}{\gamma_k}.
\end{equation}
Consequently, combining \eqref{Eq.MGFgamma-k-b} with \eqref{Eq.e2eSNR-2} in conjunction with the fact that all $\gamma_k$, $k=1, 2, \cdots, K$, are independent, we get the MGF of $1/\gamma_{e2e}$, given by
\begin{equation} \label{Eq.MGFe2eSNR-1}
M_{\frac{1}{\gamma_{e2e}}}(s) = \prod_{k=1}^{K}\Psi\left(1, 0; \frac{s}{a_k}\right).
\end{equation}

With the MGF of $1/\gamma_{e2e}$ developed, we finally obtain the CDF and MGF of $\gamma_{e2e}$. More specifically, the CDF of $\gamma_{e2e}$ can be derived as follows:
\begin{eqnarray}
F_{\gamma_{e2e}}(\gamma)
&  =  &1-F_{\frac{1}{\gamma_{e2e}}}\left(\frac{1}{\gamma}\right) \nonumber \\
&  =  &1-\left.\mathcal{L}^{-1}\left\{\frac{1}{s}\prod_{k=1}^{K}\Psi\left(1, 0; \frac{s}{a_k}\right)\right\}\right\vert_{s=\frac{1}{\gamma}}, \label{Eq.CDFe2eSNR}
\end{eqnarray}
where the operator $\mathcal{L}^{-1}\left\{.\right\}$means the inverse Laplace transform and $f(x)\left\vert_{x=a}\right.$ means the value of $f(x)$ at $x=a$. On the other hand, by virtue of \cite[Eq.(7)]{Asghari10} and performing some algebraic manipulations, the MGF of $\gamma_{e2e}$ can be shown as
\begin{equation}  \label{Eq.MGFe2eSNR}
\mathrm{M}_{\gamma_{e2e}}(s)
= 1-2\sqrt{s}\int_0^\infty{J_1\left(2x\sqrt{s}\right)\prod_{k=1}^{K}\Psi\left(1, 0; \frac{x^2}{a_k}\right)}\,\mathrm{d}x,
\end{equation}
where $J_1(x)$ denotes the first-order Bessel function of the first kind \cite[Eq.(8.402)]{Gradshteyn07}.

In general, when the number of hops $K>2$, the CDF in \eqref{Eq.CDFe2eSNR} and the MGF in \eqref{Eq.MGFe2eSNR} of the end-to-end SNR $\gamma_{e2e}$ cannot be expressed in closed-form. In particular, when $K=2$, multi-hop relaying reduces to dual-hop relaying and the CDF and MGF of $\gamma_{e2e}$ can be found in closed-form as follows.

\emph{Special Cases}: When $K=2$, the CDF in \eqref{Eq.CDFe2eSNR} reduces to
\begin{eqnarray}
F_{\gamma_{e2e}}(\gamma)
&  =  &1-\left.\mathcal{L}^{-1}\left\{\frac{1}{s}\Psi\left(1, 0; \frac{s}{a_1}\right)\Psi\left(1, 0; \frac{s}{a_2}\right)\right\}\right\vert_{s=\frac{1}{\gamma}} \nonumber \\
&  =  &1- \frac{1}{2}\left(1+\frac{\gamma}{a_1}\right)^{-1}\left(1+\frac{\gamma}{a_2}\right)^{-1}  \nonumber \\
&      &{}\times {_2F_1}\left(1, 1; 3; \frac{1+\left(\frac{1}{a_1}+\frac{1}{a_2}\right)\gamma}{\left(1+\frac{\gamma}{a_1}\right)\left(1+\frac{\gamma}{a_2}\right)}\right),
\label{Eq.CDFe2eSNR-2}
\end{eqnarray}
where \cite[vol.5, Eq.(3.34.6.3)]{Prudnikov86} was exploited to derive \eqref{Eq.CDFe2eSNR-2} and $_2F_1(a,\,b;\,c\,;x)$ is the Gaussian hypergeometric function \cite[Eq.(9.100)]{Gradshteyn07}.

\newcounter{mytempeqncnt1}
\begin{figure*}[!t]
\normalsize
\setcounter{mytempeqncnt1}{\value{equation}}
\setcounter{equation}{21}
\begin{equation}  \label{Eq.MGFe2eSNR-2}
\mathrm{M}_{\gamma_{e2e}}(s)
= 1-2\sqrt{s}\int_0^\infty{J_1\left(2x\sqrt{s}\right)\Psi\left(1, 0; \frac{x^2}{a_1}\right)\Psi\left(1, 0; \frac{x^2}{a_2}\right)}\,\mathrm{d}x.
\end{equation}
\setcounter{equation}{\value{mytempeqncnt1}}
\hrulefill
\vspace*{4pt}
\end{figure*}
\setcounter{equation}{22}

On the other hand, the MGF in \eqref{Eq.MGFe2eSNR} reduces to \eqref{Eq.MGFe2eSNR-2} shown at the top of the next page. Then, by expressing the confluent hypergeometric function $\Psi(1, 0; x)$ in the integrand of \eqref{Eq.MGFe2eSNR-2} in terms of Fox's $H$-function and performing Hankel transform over the product of two $H$-functions (see Appendix~\ref{Appendix-B}), \eqref{Eq.MGFe2eSNR-2} can be expressed according to the following closed-form (for detailed derivations, please refer to Appendix~\ref{Appendix-C}):
\begin{equation}  \label{Eq.MGFe2eSNR-3}
\mathrm{M}_{\gamma_{e2e}}(s)
= 1-
G_{2,\,[1:\,1],\,0,\,[2:\,2]}^{1,\,1,\,1,\,2,\,2}
\left[\begin{gathered}\frac{1}{a_1\,s}\\ \\ \frac{1}{a_2\,s}\end{gathered}
\left\vert \begin{gathered} 1,\, 0 \\ 0;\,0 \\ - \\ 0, 1;\,0, 1\end{gathered}\right.\right],
\end{equation}
where $G[^x_y \,\vert\,.\,]$ denotes the generalized Meijer's $G$-function of two variables defined in \cite[Appendix~A]{{XiaTWC11Oct}}. Although there is no built-in function to compute the above bivariate $G$-function in popular mathematical softwares such as Matlab and Mathematica, an efficient Mathematica algorithm was recently developed in \cite{Ansari11}. Hence, the exact analytical results of the CDF in \eqref{Eq.CDFe2eSNR-2} and the MGF in \eqref{Eq.MGFe2eSNR-3} with respect to $K=2$ can be used to serve as a benchmark for the numerical results of \eqref{Eq.CDFe2eSNR} and \eqref{Eq.MGFe2eSNR} with respect to the general cases with $K>2$.

Although the exact CDF and MGF of the end-to-end SNR were developed as above, they are a bit on the complex side for further processing. In order to proceed and in particular to gain illuminating insights into system performance, in the next subsection we propose a pair of lower and upper bounds on the end-to-end SNR and study its limiting distribution functions, by using the extreme value theory.

\subsection{Limiting Distribution Functions of the Lower and Upper Bounds on the End-to-End SNR}
\label{Subsection:LimitingDistribution}

By virtue of \eqref{Eq.e2eSNR}, the end-to-end SNR $\gamma_{e2e}$ can be upper bounded by
\begin{equation} \label{Eq.e2eSNRUpperBound}
\gamma_{e2e} \le \left(\max_{k=1, \cdots, K}\frac{1}{\gamma_k}\right)^{-1} = \min_{k=1, \cdots, K}{\gamma_k}.
\end{equation}
On the other hand, it is clear that $\gamma_{e2e}$ is lower bounded by
\begin{equation} \label{Eq.e2eSNRLowerBound}
\gamma_{e2e} \ge \left(K \max_{k=1, \cdots, K}\frac{1}{\gamma_k}\right)^{-1} = \frac{1}{K}\min_{k=1, \cdots, K}{\gamma_k}.
\end{equation}
Consequently, combining \eqref{Eq.e2eSNRUpperBound} and \eqref{Eq.e2eSNRLowerBound} yields the lower and upper bounds on $\gamma_{e2e}$, namely,
\begin{equation} \label{Eq.e2eSNRBounds}
\underbrace{\frac{1}{K}\min_{k=1, \cdots, K}{\gamma_k}}_{\gamma_{e2e}^\mathrm{lower}}
\le \gamma_{e2e}
\le  \underbrace{\min_{k=1, \cdots, K}{\gamma_k}}_{\gamma_{e2e}^\mathrm{upper}}.
\end{equation}
Moreover, it is observed from \eqref{Eq.e2eSNRBounds} that
\begin{equation} \label{Eq.e2eSNRBoundsRelation}
\gamma_{e2e}^\mathrm{lower}
= \frac{1}{K}\,\gamma_{e2e}^\mathrm{upper}.
\end{equation}

Now, we derive the distribution function of $\gamma_{e2e}^\mathrm{upper}$. That of $\gamma_{e2e}^\mathrm{lower}$ follows in a straightforward manner. To this end, integrating the PDF of $\gamma_k$ given by \eqref{Eq.PDFgamma-k} with respect to $\gamma$ yields its CDF:
\begin{equation} \label{Eq.CDFgamma-k}
F_{\gamma_k}(\gamma) = 1 - \frac{a_k}{\gamma+a_k}.
\end{equation}
Subsequently, recalling the theory of order statistics, we obtain the CDF of $\gamma_{e2e}^\mathrm{upper}$, given by
\begin{equation}
F_{\gamma_{e2e}^\mathrm{upper}}(\gamma)
= 1-\prod_{k=1}^K{\left(\frac{a_k}{\gamma+a_k}\right)}.   \label{Eq.CDFUpperBound}
\end{equation}
Furthermore, combining \eqref{Eq.e2eSNRBoundsRelation} with \eqref{Eq.CDFUpperBound}, we obtain the CDF of $\gamma_{e2e}^\mathrm{lower}$, that is
\begin{equation}\label{Eq.CDFLowerBound}
F_{\gamma_{e2e}^\mathrm{lower}}(\gamma)
= 1-\prod_{k=1}^K{\left(\frac{a_k}{K\gamma+a_k}\right)}.
\end{equation}

Although \eqref{Eq.CDFUpperBound} and \eqref{Eq.CDFLowerBound} can be used to measure system performance such as the outage probability of the end-to-end SNR, they are too complex to offer any insight into system performance. Therefore, in the following we derive the limiting distribution of $\gamma_{e2e}^\mathrm{upper}$ and $\gamma_{e2e}^\mathrm{lower}$ as the number of hops $K \to \infty$.

First of all, it is assumed that all channel fast-fading coefficients in the considered system are independently and identically distributed (i.i.d.) and AWGNs at all secondary nodes along the multi-hop relaying link have the same variance (i.e. $\sigma_1^2 = \cdots = \sigma_K^2 \triangleq \sigma^2$). Moreover, it is assumed that the path-loss ratios $\eta_1 = \cdots = \eta_K \triangleq \eta$. We note that the latter assumption is feasible since, by recalling the definition of $\eta_k = \left({l_k}/{d_k}\right)^\epsilon$ given by \eqref{Eq.NormalizedDistance}, larger $l_k$ (i.e. larger distance between secondary user $\mathrm{SU}_{k-1}$ and primary receiver $\mathrm{PU}_1$) implies higher transmit power at $\mathrm{SU}_{k-1}$ and in turn allows larger transmission distance $d_k$ (i.e. larger distance between consecutive secondary users $\mathrm{SU}_{k-1}$ and $\mathrm{SU}_{k}$); thus, the ratio of $l_k$ to $d_k$, or equivalently $\eta_k$, can be fixed $\forall k \in [1, K]$. Accordingly, the parameter $a_k$ in \eqref{Eq.CDFgamma-k} keeps constant (i.e. $a_1 = \cdots = a_K \triangleq a$) and the received SNRs ($\gamma_k$, $k=1, 2, \cdots, K$) at consecutive hops are i.i.d. random variables. More specifically, the CDF of $\gamma_k$ in \eqref{Eq.CDFgamma-k} reduces to
\begin{equation} \label{Eq.CDFgamma-k-IID}
F_{\gamma_k}(\gamma) = 1 - \frac{a}{\gamma+a},
\end{equation}
where $a = \lambda\eta/\sigma^2$.

Based on \eqref{Eq.CDFgamma-k-IID} and using the extreme value theory \cite{Galambos87}, we can obtain the limiting distribution of the upper bound on the end-to-end SNR, which is summarized in the following theorem.
\begin{theorem}  \label{Theorem1}
Assuming that the received SNRs at $K$ consecutive hops are i.i.d., according to \eqref{Eq.CDFgamma-k-IID}, the limiting CDF and PDF of the upper bound \eqref{Eq.e2eSNRBounds} on the end-to-end SNR are given by
\begin{equation}  \label{Eq.e2eSNRUpperBoundCDF}
\lim\limits_{K \to \infty}{F_{\gamma_{e2e}^\mathrm{upper}}\left(\frac{\lambda \eta}{(K-1)\sigma^2}\,\gamma\right)}
= 1-\exp\left(-\gamma\right)
\end{equation}
and
\begin{equation} \label{Eq.e2eSNRUpperBoundPDF}
\lim\limits_{K \to \infty}{\frac{\lambda \eta}{(K-1)\sigma^2}\,f_{\gamma_{e2e}^\mathrm{upper}}\left(\frac{\lambda \eta}{(K-1)\sigma^2}\,\gamma\right)}
= \exp\left(-\gamma\right),
\end{equation}
respectively.
\begin{IEEEproof}
See Appendix~\ref{Appendix-D}.
\end{IEEEproof}
\end{theorem}

By using the relation \eqref{Eq.e2eSNRBoundsRelation} along with \eqref{Eq.e2eSNRUpperBoundCDF}-\eqref{Eq.e2eSNRUpperBoundPDF}, the limiting CDF and PDF of the lower bound on the end-to-end SNR can be derived in a straightforward manner. In summary, the results are presented in the following theorem.
\begin{theorem}  \label{Theorem2}
The limiting CDF and PDF of the lower bound \eqref{Eq.e2eSNRBounds} on the end-to-end SNR are given by
\begin{equation}  \label{Eq.e2eSNRLowerBoundCDF}
\lim\limits_{K \to \infty}{F_{\gamma_{e2e}^\mathrm{lower}}\left(\frac{\lambda \eta}{K(K-1)\sigma^2}\,\gamma\right)}
= 1-\exp\left(-\gamma\right)
\end{equation}
and
\begin{equation}\label{Eq.e2eSNRLowerBoundPDF}
\lim\limits_{K \to \infty}{\frac{\lambda \eta}{K(K-1)\sigma^2}\,f_{\gamma_{e2e}^\mathrm{lower}}\left(\frac{\lambda \eta}{K(K-1)\sigma^2}\,\gamma\right)}
= \exp\left(-\gamma\right),
\end{equation}
respectively.
\end{theorem}

It is observed from \eqref{Eq.e2eSNRUpperBoundCDF}-\eqref{Eq.e2eSNRLowerBoundPDF} that, after appropriate normalization, the limiting distribution of either the upper or the lower bound on the end-to-end SNR is of the standard exponential distribution. More importantly, \eqref{Eq.e2eSNRUpperBoundCDF}-\eqref{Eq.e2eSNRLowerBoundPDF} capture the relationship between the end-to-end SNR and the number of relaying hops ($K$), the power-allocation parameter ($\lambda$), the path-loss ratio ($\eta$) and the noise variance ($\sigma^2$), in an explicit and simple way. For example, it is evident from \eqref{Eq.e2eSNRUpperBoundCDF} that larger $K$ and $\sigma^2$ will deteriorate the outage probability for a fixed outage threshold value, whereas larger $\lambda$ or $\eta$ will improve the outage probability. In the next section, the resultant \eqref{Eq.e2eSNRUpperBoundCDF}-\eqref{Eq.e2eSNRLowerBoundPDF} are applied to analyze and gain insights into the system performance.

Notice that, although the extreme value theory used in the proof of \emph{Theorem 1} assumes that the number of relaying hops $K$ is sufficiently large, thanks to the rapid convergence speed of the resultant limiting functions, Eqs. \eqref{Eq.e2eSNRUpperBoundCDF}-\eqref{Eq.e2eSNRLowerBoundPDF} can apply to the cases even if the value of $K$ is small or moderate. Actually, the extreme value theory was already exploited to analyze system performance of wireless communications, see e.g. \cite{Xia09TVT02, Du13TSP} with respect to multi-input multi-output (MIMO) systems.

More specifically, since the parent distribution function \eqref{Eq.CDFgamma-k-IID} of the considered order statistics is of the generalized Pareto distribution, its speed of convergence in the context of extreme value theory is at least of algebraic order, namely, the error committed by the replacement of the exact distribution of the extremes by their limiting forms in \eqref{Eq.e2eSNRUpperBoundCDF}-\eqref{Eq.e2eSNRLowerBoundPDF} is at least of the order of $1/K$ \cite[Theorem 2.2.5]{Folk10}. On the other hand, the effectiveness of the obtained limiting distributions with respect to practical values of $K = 4, 8$ is illustrated in the following Fig.~\ref{Fig.OutageInterferenceHops}, compared to the simulation results. Also, for comparison purposes, the exact numerical results of \eqref{Eq.CDFe2eSNR-2} and the simulation results pertaining to the case with $K = 2$ are presented in Fig.~\ref{Fig.OutageInterferenceHops}. For more theoretical details on the convergence of limiting distribution of extreme order statistics, we refer the interested reader to \cite[Section 2.10]{Galambos87}.

\textbf{Remark 5 (I.I.D. Channels):}
For ease of mathematical tractability of the above theorems, it is assumed that the variances of AWGNs at all nodes are identical in Section IV-B. In general, however, AWGNs at different nodes may have different variances and different channels may have different average power gain values. Actually, for ease of theoretical analysis, the variance of AWGN and the average value of channel power gain can be absorbed without loss of generality into a single parameter, namely, average SNR. For better clarity, let us take a basic transmitter-receiver communication over fading channels for instance. Specifically, the instantaneous received SNR per symbol is given by $\gamma = \frac{{E_s}}{\sigma^2}\,h$, where $E_s$ denotes the symbol energy at the transmitter, $h$ refers to the instantaneous channel gain, and $\sigma^2$ stands for the variance of the AWGN at the receiver. Clearly, the average received SNR is $\mathcal{E}\left\{\gamma\right\} =  \frac{E_s}{\sigma^2}\,\mathcal{E}\left\{h\right\}\triangleq \bar{\gamma}$. Then, by combining the above two equations, the instantaneous received SNR can be reformulated as $\gamma = \bar{\gamma}\, \frac{h}{\mathcal{E}\left\{h\right\}}/1 =  \bar{\gamma}g/1$, where $g \triangleq \frac{h}{\mathcal{E}\left\{h\right\}}$ denotes the normalized channel gain and the unity in the denominator stands for the normalized noise variance. In other words, the average SNR $\bar{\gamma}$ captures the effects of both the mean of the channel gain and the noise variance. Accordingly, changing the value of $\bar{\gamma}$ in simulation setting is equivalent to changing both the mean gain and the noise variance. Actually, similar assumption of i.i.d. fading channels is widely used in the open literature, see e.g.  \cite{Hong12TCOMJan}.

\section{System Performance Analysis}
\label{Section:PerformanceAnalysis}
With the obtained distribution functions of the lower and upper bounds on the end-to-end SNR, in this section we investigate the end-to-end performance of spectrum-sharing multi-hop AF relaying, in terms of the outage probability, the diversity and coding gains, and the achievable data rate.

\subsection{Simulation Scenario and Parameter Setting}

\begin{figure}[t]
\centering
\includegraphics [width=2.5in, clip, keepaspectratio]{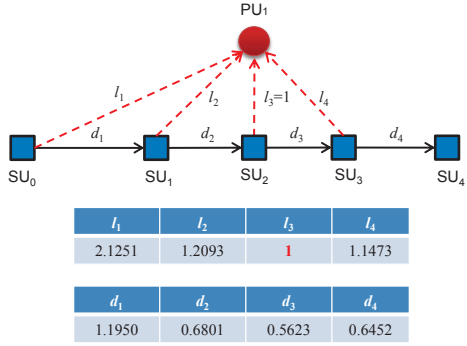}
\caption{Simulation scenario and distance parameters, where the number of relaying hops $K=4$ and the path-loss ratio $\eta = 10$.}
\label{Fig.SimulationSetting}
\end{figure}

In the simulation experiments, the number of hops is set to an even integer ($K = 2$, $4$, or $8$) and all secondary nodes ($\mathrm{SU}_0, \cdots, \mathrm{SU}_K$) are deployed sequentially along a straight line. Also, the multi-hop link is perpendicular to the interference link from the middle relaying node, i.e. $\mathrm{SU}_\frac{K}{2}$, to the primary receiver $\mathrm{PU}_1$. Further, in order to determine the geometry of all nodes, the distance between $\mathrm{SU}_\frac{K}{2}$ and $\mathrm{PU}_1$ is normalized to unity (i.e. $l_{\frac{K}{2}+1} \equiv 1$), and the large-scale path-loss ratio \eqref{Eq.NormalizedDistance} is set to $\eta = \eta_k \triangleq d_k^{-\epsilon}/l_k^{-\epsilon} = 10$ with the path-loss exponent $\epsilon = 4$, $\forall k \in [1, K]$. With these definitions in mind, the distance parameters used to locate each relaying node can be easily determined. For illustration purposes, Fig.~\ref{Fig.SimulationSetting} shows a typical simulation scenario with $K = 4$ hops, where the $4$-hop link is perpendicular to the interference link from $\mathrm{SU}_2$ to $\mathrm{PU}_1$. Since the distance between $\mathrm{SU}_2$ and $\mathrm{PU}_1$ is normalized to unity, i.e. $l_3 = 1$, the distance between $\mathrm{SU}_2$ and $\mathrm{SU}_3$ can be computed by $d_3 = l_3 \times \eta^{-\frac{1}{\epsilon}} = 1 \times 10^{-\frac{1}{4}} = 0.5623$. Then, the distance from $\mathrm{SU}_3$ to $\mathrm{PU}_1$ can be computed by $l_4 = \sqrt{l_3^2+d_3^2} = \sqrt{1^2+0.5623^2} = 1.1473$. Other distance parameters can be found in a similar way and are as shown in Fig.~\ref{Fig.SimulationSetting}.

In the ensuing Monte-Carlo simulations, the gain of channel fast-fading is subject to Rayleigh distribution with unit mean, and the variance of AWGNs at all nodes is set to unity. Furthermore, the energy of each transmitted symbol is set to unity but is scaled before transmission by a constant corresponding to the average SNR. The power-allocation parameter at each secondary node along the multi-hop link is off-line computed as per (\ref{Eq.Lambda}). Moreover, for each channel realization, the optimal transmit power at each secondary node is computed according to (\ref{Eq.S1CapacityC}).

\subsection{Outage Probability}
\label{Subsection:OutageProbability}
As an important performance indicator, outage probability is defined as the probability that the instantaneous end-to-end SNR falls below a prescribed threshold value~$\gamma_{_\mathrm{th}}$. Furthermore, since outage probability is a monotonically decreasing function of the end-to-end SNR, in light of \eqref{Eq.e2eSNRUpperBoundCDF} and \eqref{Eq.e2eSNRLowerBoundCDF}, the outage probability of the system under study can be shown to be bounded by
\begin{equation} \label{Eq.OutageProbability}
F_{\gamma_{e2e}^\mathrm{upper}}\left(\gamma_{_\mathrm{th}}\right)
\le\mathrm{P_{outage}}(\gamma_{_\mathrm{th}})
\le F_{\gamma_{e2e}^\mathrm{lower}}\left(\gamma_{_\mathrm{th}}\right).
\end{equation}

In order to confirm the effectiveness of the preceding analysis, extensive simulation experiments are performed to compare the simulation results of outage probability with the numerical results of the above lower and upper bounds.

\begin{figure}[t]
\centering
\includegraphics [width=3.0in, clip, keepaspectratio]{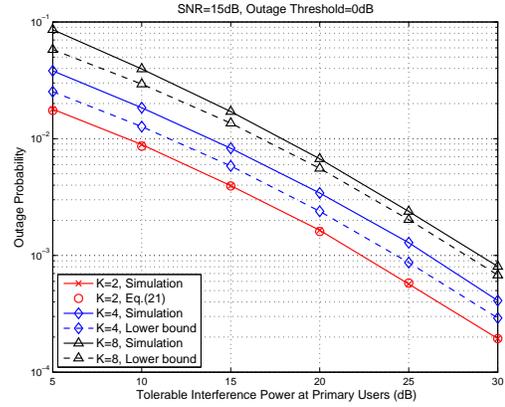}
\caption{Outage probability versus tolerable interference power for different numbers of relaying hops $K$.}
\label{Fig.OutageInterferenceHops}
\end{figure}

Figure~\ref{Fig.OutageInterferenceHops} shows the effect of the number of relaying hops on the outage probability, where the outage probability versus the tolerable interference power at PUs is plotted with the number of hops varying from $2$ to $8$. In particular, when the number of relaying hops $K=2$, the numerical results of outage probability were computed by using the analytical CDF in \eqref{Eq.CDFe2eSNR-2}. In such a case, it is seen from Fig.~\ref{Fig.OutageInterferenceHops} that the analytical results agree with the simulation results very well. On the other hand, when the value of $K$ increases from $4$ to $8$, it is observed that the simulation results of outage probability become more and more tight with the lower bound (computed by \eqref{Eq.OutageProbability} in conjunction with \eqref{Eq.e2eSNRUpperBoundCDF}), especially in the medium to high tolerable-interference-power regions. These observations demonstrate the effectiveness of the previous limiting performance analysis even when the number of relaying hops is small.

\begin{figure}[t]
\centering
\includegraphics [width=3.0in, clip, keepaspectratio]{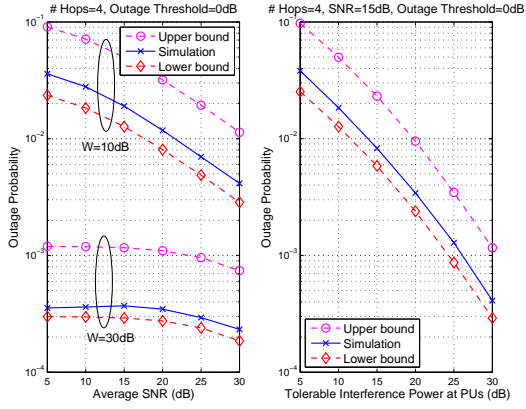}
\caption{Outage probability of $4$-hop spectrum-sharing AF relaying.}
\label{Fig.Outage4Hop}
\end{figure}

Figure~\ref{Fig.Outage4Hop} depicts the outage probability of $4$-hop AF relaying in spectrum-sharing environment. In particular, the left sub-plot shows the outage probability versus the average SNR with respect to different prescribed tolerable interference powers at PUs, namely, $W=10~\rm{and}~30$dB. It is observed that the outage probability declines slowly when $W=10$dB and it almost keeps constant when $W=30$dB over a wide range of average SNR of interest. On the other hand, the right sub-plot illustrates the outage probability versus the tolerable interference power at PUs with $\mathrm{SNR}=15$dB. It is seen that the outage probability decreases sharply with the tolerable interference power. Furthermore, both sub-plots of Fig.~\ref{Fig.Outage4Hop} show that the lower bound is tighter with the simulation results than the upper bound. This is because when the received SNRs at consecutive hops are i.i.d., the end-to-end SNR in \eqref{Eq.e2eSNRBounds} approaches its upper bound. In other words, the outage probability of multi-hop AF relaying under spectrum-sharing constraint is dominated by the tolerable interference power at the primary receiver and it is highly predictable by the lower bound given by \eqref{Eq.OutageProbability} as well as \eqref{Eq.e2eSNRUpperBoundCDF}.

In the next subsection, we investigate the fundamental reason why different parameters (including the number of relaying hops, the noise variance, the water level of water-filling power allocation, and the path-loss ratio) have different effects on outage probability.

\subsection{Diversity and Coding Gains}
\label{Subsection:DiversityCodingGains}

Now, we derive the diversity gain and the coding gain of the system under study, which are critical for indicating outage probability and/or symbol error probability at high SNR. According to \cite{WangTCOM03Aug}, if the limit of the PDF of the end-to-end SNR $\gamma_{e2e}$ can be expressed as:
\begin{equation}   \label{Eq.WangI}
\lim\limits_{\gamma \to 0}f_{\gamma_{e2e}}(\gamma) = b\gamma^t+\emph{o}\left(\gamma^{t+\epsilon}\right),
\end{equation}
where $b, \epsilon > 0$ and $\emph{o}(\,.\,)$ pertain to the Landau notation, then the diversity gain and the coding gain are given by
\begin{equation}  \label{Eq.WangII}
G_d = t+1
\end{equation}
and
\begin{equation}  \label{Eq.WangIII}
G_c = p\left(\frac{2^{t}\,b\,\Gamma(t+\frac{3}{2})}{\sqrt{\pi}(t+1)}\right)^{-\frac{1}{t+1}},
\end{equation}
respectively, where $p$ is a positive constant relevant to the used constellation, for example, $p=2$ for BPSK.

Substituting \eqref{Eq.e2eSNRUpperBoundPDF} into \eqref{Eq.WangI} yields
\begin{eqnarray}
\lim\limits_{\gamma \to 0}f_{\gamma_{e2e}}(\gamma)
&  =  & \frac{\sigma^2}{\lambda \eta}(K-1)  \lim\limits_{\gamma \to 0}{\exp\left(-\frac{\sigma^2}{\lambda \eta}(K-1)\gamma\right)}     \nonumber\\
&  =  & \frac{\sigma^2}{\lambda \eta}(K-1)\left[1+\emph{O}\left(\gamma\right)\right]                                \label{Eq.DiversityCodingGains-2} \\
&  =  & \frac{\sigma^2}{\lambda \eta}(K-1)+\emph{O}\left(\gamma\right).                                                \label{Eq.DiversityCodingGains-3}
\end{eqnarray}
Comparing \eqref{Eq.DiversityCodingGains-3} with \eqref{Eq.WangI} results in $t=0$ and $b=\frac{\sigma^2}{\lambda \eta}(K-1)$. Afterwards, substituting them into \eqref{Eq.WangII}-\eqref{Eq.WangIII} and performing some algebraic manipulations, we obtain
\begin{equation}  \label{Eq.DiversityCodingGains-4}
G_d = 1,
\end{equation}
\begin{equation}  \label{Eq.DiversityCodingGains-5}
G_c = \frac{2p\lambda \eta}{(K-1)\sigma^{2}}.
\end{equation}

Equation~\eqref{Eq.DiversityCodingGains-4} implies that, for multi-hop AF relaying in the context of spectrum-sharing CR, the achievable diversity gain is unity regardless of the number of relaying hops. This is in agreement with the observation in Fig.~\ref{Fig.OutageInterferenceHops} where the slopes of all curves are always unity.

Equation~\eqref{Eq.DiversityCodingGains-5} demonstrates that the coding gain is proportional to the power-allocation parameter $\lambda$. This is exactly the fundamental reason why the optimal transmit-power allocation discussed in Section~\ref{Section:PowerAllocation} benefits improving system performance. Moreover, \eqref{Eq.DiversityCodingGains-5} shows that the coding gain is proportional to the path-loss ratio $\eta$. This is because, by recalling $\eta = (l/d)^\epsilon$, for a fixed distance $d$ between a secondary transmitter and its receiver, larger $\eta$, i.e. larger distance $l$ between the secondary transmitter and the primary receiver, allows higher transmit power at the secondary transmitter. On the other hand, \eqref{Eq.DiversityCodingGains-5} also indicates that the coding gain is inversely proportional to the product of $K-1$ and $\sigma^2$. That is, larger number of intermediate relaying nodes ($K-1$) and larger value of noise variance ($\sigma^2$) will deteriorate the coding gain. This is widely known as the effect of noise accumulation in AF relaying networks \cite{BoradeTIT07Oct}. The relationship between the coding gain and the number of hops is in accordance with the observation in Fig.~\ref{Fig.OutageInterferenceHops} where increasing $K$ decreases the coding gain and the curves of outage probability versus the average SNR shift to the right.

In summary, we conclude that, for the spectrum-sharing based multi-hop AF relaying transmission, the diversity gain is always unity regardless of the number of relaying hops. On the other hand, the coding gain is proportional to the water level ($\lambda$) of the water-filling power allocation at secondary transmitter and the path-loss ratio ($\eta$) at each hop, yet is inversely proportional to the accumulated noise at the destination, i.e. $(K-1)\sigma^2$.

\subsection{Average Symbol Error Probability}

It is worth pointing out that, with the MGF in \eqref{Eq.MGFe2eSNR} ($\forall K$) or \eqref{Eq.MGFe2eSNR-3} ($K=2$) of the end-to-end SNR, the average symbol error probability (ASEP) can be obtained by using the MGF-based methodology. Similarly, with the obtained distribution functions of the upper bound on the end-to-end SNR, the ASEP can be derived by using the PDF/CDF-based methodology. However, since ASEP behaves like outage probability and they are characterized by the above diversity and coding gains, we do not discuss ASEP in this paper. For the interested reader, we refer to \cite{WangTCOM03Aug, Maaref09TCOM01}.

\subsection{Achievable Data Rate}
Based on the seminal Shannon theorem, the maximum achievable data rate (in the unit of bit/s/Hz) of secondary destination is given by
\begin{eqnarray}
R
&  =  &  \frac{1}{K}\int_0^\infty{\log_2(1+\gamma)\,f_{\gamma_{e2e}}(\gamma)}\,\mathrm{d}\gamma                                       \label{Eq.DataRate-1} \\
& \le  &  \frac{1}{K}\int_0^\infty{\log_2(1+\gamma)\,f_{\gamma_{e2e}^{\mathrm{upper}}}(\gamma)}\,\mathrm{d}\gamma         \label{Eq.DataRate-2} \\
&  =  &  \frac{(K-1)\sigma^2}{\lambda \eta K \ln{2}}   \nonumber \\
&      &{}\times \int_0^\infty{\ln(1+\gamma)\,\exp\left(-\frac{(K-1)\sigma^2}{\lambda \eta}\gamma\right)}\,\mathrm{d}\gamma \label{Eq.DataRate-3}\\
&  =  &  \frac{1}{K\ln{2}}\exp\left(\frac{(K-1)\sigma^2}{\lambda \eta}\right)\mathrm{E_1}\left(\frac{(K-1)\sigma^2}{\lambda \eta}\right)  \label{Eq.DataRate-4}
\end{eqnarray}
where the coefficient $1/K$ in \eqref{Eq.DataRate-1} was introduced since $K$ consecutive time slots are involved to complete a single data transmission from a secondary source to its final destination, \eqref{Eq.e2eSNRUpperBoundPDF} was substituted into \eqref{Eq.DataRate-2} to reach \eqref{Eq.DataRate-3}, and \cite[vol.1, Eq.(2.6.23.5)]{Prudnikov86} was exploited to derive \eqref{Eq.DataRate-4}, with $\mathrm{E_1}(x) = \int_x^\infty{\frac{e^{-t}}{t}}\,\mathrm{d}t$, $x > 0$ being the exponential integral function.

\begin{figure}[t]
\centering
\includegraphics [width=3.0in, clip, keepaspectratio]{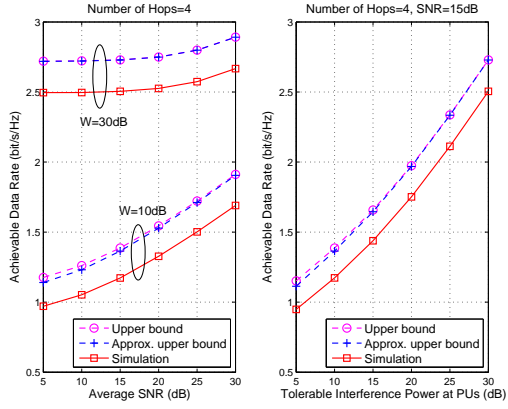}
\caption{Achievable data rate of $4$-hop spectrum-sharing AF relaying.}
\label{Fig.Rate4Hop}
\end{figure}

Moreover, it is remarkable that, when the average SNR or the tolerable interference power at PUs is medium or high, the term $\frac{(K-1)\sigma^2}{\lambda \eta} \ll 1$ holds. In such a case, \eqref{Eq.DataRate-4} reduces to
\begin{eqnarray}
R
&  \lessapprox  &  \frac{1}{K\ln{2}}\left(1+\frac{(K-1)\sigma^2}{\lambda \eta}\right)                                           \nonumber \\
&                     & \times \left[-\Upsilon-\ln\left(\frac{(K-1)\sigma^2}{\lambda \eta}\right)\right]                               \label{Eq.DataRate-5} \\
&     \approx    & \frac{1}{K\ln{2}}\left[-\Upsilon-\ln\left(\frac{(K-1)\sigma^2}{\lambda \eta}\right)\right]              \label{Eq.DataRate-6} \\
&         =         & \frac{1}{K\ln{2}}\left[-\Upsilon+\ln\left(\frac{\lambda \eta}{(K-1)\sigma^{2}}\right)\right],           \label{Eq.DataRate-7}
\end{eqnarray}
where the approximations $e^x \approx 1+x$ and $\mathrm{E_1}(x) \approx -\Upsilon-\ln{x}$, $x \ll 1$ with $\Upsilon = 0.5772 \cdots $ being the Euler's constant were exploited to derive \eqref{Eq.DataRate-5}. Equation \eqref{Eq.DataRate-7} establishes the vital link between the achievable data rate and the average received SNR (i.e. $\frac{\lambda \eta}{(K-1)\sigma^{2}}$) of the destination node in the spectrum-sharing based multi-hop relaying system under study, in a very simple and explicit manner. It is clear from \eqref{Eq.DataRate-7} that higher water level (i.e. $\lambda$) of the water-filling power allocation and larger path-loss ratio (i.e. $\eta$) at each relaying node improves the achievable data rate whereas an increase in the noise accumulation (i.e. $(K-1)\sigma^{2}$) deteriorates the data rate.

Figure~\ref{Fig.Rate4Hop} depicts the achievable data rate of $4$-hop AF relaying in spectrum-sharing environment, where the simulation settings are identical to those used for Fig.~\ref{Fig.Outage4Hop}. The left sub-plot shows that the achievable data rate improves slightly with increasing average SNR whereas the right sub-plot illustrates that the achievable data rate increases sharply with increasing tolerable interference power at the primary receiver. On the other hand, both sub-plots of Fig.~\ref{Fig.Rate4Hop} show that the upper bound computed by \eqref{Eq.DataRate-4} is tight with the simulation results, and the approximated upper bound computed by \eqref{Eq.DataRate-7} works very well in the whole SNR and/or tolerable interference power regions of interest. In particular, the closed-form upper bound \eqref{Eq.DataRate-4} perfectly reflects the growing trend in the achievable data rate.

When the number of hops $K=2$, the achievable data rate can be accurately obtained in a numerical way. Specifically, by using the integration-by-parts method, \eqref{Eq.DataRate-1} can be rewritten as
\begin{equation} \label{Eq.DataRate-8}
R = \frac{1}{2\ln{2}}\int_0^\infty\frac{1}{1+\gamma}\left[1-F_{\gamma_{e2e}}(\gamma)\right]\,\mathrm{d}\gamma.
\end{equation}
Hence, inserting \eqref{Eq.CDFe2eSNR-2} into \eqref{Eq.DataRate-8} yields the achievable data rate at the secondary destination following $K=2$ hops, that is
\begin{eqnarray} \label{Eq.DataRate-9}
R
&  =  & \frac{1}{4\ln{2}}\int_0^\infty\frac{1}{1+\gamma}\left(1+\frac{\gamma}{a_1}\right)^{-1}\left(1+\frac{\gamma}{a_2}\right)^{-1} \nonumber \\
&      & {}\times{_2F_1}\left(1, 1; 3; \frac{1+\left(\frac{1}{a_1}+\frac{1}{a_2}\right)\gamma}{\left(1+\frac{\gamma}{a_1}\right)\left(1+\frac{\gamma}{a_2}\right)}\right)
\,\mathrm{d}\gamma.
\end{eqnarray}
Although a closed-form expression for \eqref{Eq.DataRate-9} cannot be obtained in a straightforward manner, \eqref{Eq.DataRate-9} can be easily evaluated in a numerical way since the Gaussian hypergeometric function ${_2F_1}(1, 1; 3; x)$ in the integrand of \eqref{Eq.DataRate-9} is a built-in function in popular mathematical softwares, such as Mathematica.

\begin{figure}[t]
\centering
\includegraphics [width=3.0in, clip, keepaspectratio]{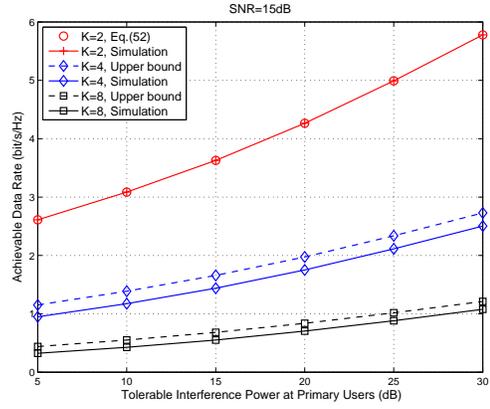}
\caption{Achievable data rate versus tolerable interference power at PUs for different numbers of relaying hops $K$.}
\label{Fig.RateInterferenceVariedHops}
\end{figure}

Figure~\ref{Fig.RateInterferenceVariedHops} shows the achievable data rate versus the average tolerable interference power at the primary receiver taking the number of relaying hops $K$ as a parameter. When $K=2$, the numerical results of achievable data rate are computed by \eqref{Eq.DataRate-9}, which are in perfect agreement with the simulation results. When $K>2$, the upper bound on achievable data rate is computed by the closed-form expression \eqref{Eq.DataRate-4}, which is tight compared to the simulation results. On the other hand, it is observed from Fig.~\ref{Fig.RateInterferenceVariedHops} that, with increasing number of hops, the achievable data rate decreases significantly. This is because more and more orthogonal transmission slots are needed to complete a single data transmission from a secondary source to its final destination, which is mathematically reflected by the multiplicative factor $1/K$ in \eqref{Eq.DataRate-7}.

\section{Conclusion}
\label{Section:Conclusion}
In the context of spectrum-sharing cognitive radio, in order to extend the coverage area of secondary users (SUs) and reduce their interference region on primary users (PUs), multi-hop AF relaying was exploited and its performance was analytically studied in this paper. Since the SUs have to satisfy the prescribed tolerable interference-power constraint imposed by PUs, their transmit power was dynamically allocated and optimally determined. By analyzing the end-to-end system performance in terms of the outage probability, the diversity and coding gains and the achievable data rate, the dominant factors on the system performance were explicitly identified. In particular, the relationship between diversity/coding gains and the number of intermediate relays, the water level of the optimal water-filling power allocation, the path-loss ratio of the desired link to the interference link at each hop and the noise variance, was explicitly established.

\appendices

\section{Derivation of Eq. \eqref{Eq.PDFgamma-k}}
\label{Appendix-A}
To derive the PDF of $\gamma_k$ defined by \eqref{Eq.gamma2PA}, we start from its CDF. By definition, the CDF of $\gamma_k$ is given by
\begin{eqnarray}
F_{\gamma_{_k}}\left(\gamma\right)
&  =  & \mathrm{Pr}\left\{\gamma_k = 0\right\} + \mathrm{Pr}\left\{\gamma_k < \gamma | \gamma_k >0\right\}   \nonumber \\
&  =  & \mathrm{Pr}\left\{\gamma_k = 0\right\} + \frac{\mathrm{Pr}\left\{0 < \gamma_k < \gamma \right\}}{\mathrm{Pr}\left\{\gamma_k >0\right\}}   \nonumber \\
&  =  & F_{V_2}\left(\frac{\sigma_k^2}{\lambda_{k-1} \, \eta_k}\right)\nonumber \\
&      &{} + \frac{F_{V_2}\left(\frac{\sigma_k^2}{\lambda_{k-1} \, \eta_k}(\gamma+1)\right) - F_{V_2}\left(\frac{\sigma_k^2}{\lambda_{k-1} \, \eta_k}\right)}
{1-F_{V_2}\left(\frac{\sigma_k^2}{\lambda_{k-1} \, \eta_k}\right)}.    \nonumber
\end{eqnarray}
Then, taking the derivative of the above equation with respect to $\gamma$ yields the PDF of $\gamma_k$:
\begin{equation}  \label{Appendix-A-1}
f_{\gamma_{_k}}\left(\gamma\right)
=  \frac{\frac{\sigma_k^2}{\lambda_{k-1} \, \eta_k}f_{V_2}\left(\frac{\sigma_k^2}{\lambda_{k-1} \, \eta_k}(\gamma+1)\right)}
{1-F_{V_2}\left(\frac{\sigma_k^2}{\lambda_{k-1} \, \eta_k}\right)}.
\end{equation}
Finally, substituting the PDF of $V_2$ given by \eqref{Eq.PDF-V2} and its corresponding CDF into \eqref{Appendix-A-1} and performing some algebraic manipulations, we attain
\begin{equation}  \label{Appendix-A-2}
f_{\gamma_{_k}}\left(\gamma\right)
=  \left(\frac{\lambda_{k-1} \, \eta_k}{\sigma_k^2}+1\right)\left(\gamma+\frac{\lambda_{k-1} \, \eta_k}{\sigma_k^2}+1\right)^{-2},
\end{equation}
which leads to the desired \eqref{Eq.PDFgamma-k}.

\section{Hankel transform of the product of two Fox's $H$-functions}
\label{Appendix-B}
\newcounter{mytempeqncnt2}
\begin{figure*}[!t]
\normalsize
\setcounter{mytempeqncnt2}{\value{equation}}
\setcounter{equation}{54}
\begin{eqnarray} \label{Eq.AppendixA-1}
\lefteqn{\int_0^\infty{x^{\rho-1}\,J_v\left(ax\right)
H_{A,\,B}^{M,\,N}\left[bx^2 \left\vert\begin{gathered} (a_A, \alpha_A) \\ (b_B, \beta_B)\end{gathered}\right.\right]
H_{C,\,D}^{M_1,\,N_1}\left[cx^2 \left\vert\begin{gathered} (c_C, \gamma_C) \\ (d_D, \delta_D)\end{gathered}\right.\right]}
\,\mathrm{d}x}    \nonumber \\
&  =  &2^{\rho-1}a^{-\rho}
H_{2,\,[A:\,C],\,0,\,[B:\,D]}^{1,\,N,\,N_1,\,M,\,M_1}
\left[\begin{gathered}\frac{4b}{a^2}\\ \\ \frac{4c}{a^2}\end{gathered}
\left\vert \begin{gathered} \left(\frac{\rho+v}{2}, 1\right),\, \left(\frac{\rho-v}{2}, 1\right) \\ (a_A, \alpha_A);\,(c_C, \gamma_C) \\ - \\ (b_B, \beta_B);\,(d_D, \delta_D)
\end{gathered}\right.\right].
\end{eqnarray}
\setcounter{equation}{\value{mytempeqncnt2}}
\hrulefill
\end{figure*}
\setcounter{equation}{55}

For completeness of the proof of Eq. \eqref{Eq.MGFe2eSNR-3}, we reproduce the Hankel transform of the product of two Fox's $H$-functions. In particular, the complex sufficient conditions are listed in detail. Specifically, the Hankel transform of the product of two Fox's $H$-functions is given by Eq. \eqref{Eq.AppendixA-1} shown at the top of the next page \cite[Eq.(2.6.3)]{Mathai78}, where
\begin{equation} \label{Eq.AppendixA-2}
a > 0,
\end{equation}
\begin{equation} \label{Eq.AppendixA-3}
\lambda_1 > 0,
\end{equation}
\begin{equation} \label{Eq.AppendixA-4}
\lambda_2 > 0,
\end{equation}
\begin{equation} \label{Eq.AppendixA-5}
|\arg{b}| < \frac{\pi \lambda_1}{2},
\end{equation}
\begin{equation} \label{Eq.AppendixA-6}
|\arg{c}| < \frac{\pi \lambda_2}{2},
\end{equation}
\begin{equation} \label{Eq.AppendixA-7}
\sum_1^A{\alpha_j}-\sum_1^B{\beta_j} \le 0,
\end{equation}
\begin{equation} \label{Eq.AppendixA-8}
\sum_1^C{\gamma_j}-\sum_1^D{\delta_j} \le 0,
\end{equation}
\begin{equation} \label{Eq.AppendixA-9}
\Re{\left[\frac{v}{2}+\min\left(\frac{b_j}{\beta_j}\right)+\min\left(\frac{d_h}{\delta_h}\right)+\frac{\rho}{2}\right]} > 0,
\end{equation}
where $j=1, \cdots, M$ and $h=1, \cdots, M_1$, and
\begin{equation} \label{Eq.AppendixA-10}
\Re{\left[\frac{\rho}{2}+\max\left(\frac{a_j-1}{\alpha_j}\right)+\max\left(\frac{c_h-1}{\gamma_h}\right)\right]} < \frac{3}{4},
\end{equation}
where $j=1, \cdots, N$ and $h=1, \cdots, N_1$.

\section{Derivation of Eq. \eqref{Eq.MGFe2eSNR-3}}
\label{Appendix-C}
In order to derive the integration given by \eqref{Eq.MGFe2eSNR-2}, we first express the confluent hypergeometric function $\Psi(1, 0; x)$ in the integrand of \eqref{Eq.MGFe2eSNR-2} in terms of Fox's $H$-function. In particular, by use of \cite[vol.3, Eqs.(8.4.46.1) \& (8.3.2.21)]{Prudnikov86}, we have
\begin{equation}
\Psi\left(1,\,0;\,x\right)
= H_{1,\,2}^{2,\,1}\left[x \left\vert\begin{gathered} (0, 1) \\(0, 1),\,(1, 1)\end{gathered}\right.\right],  \label{Tricomi2Meijer}
\end{equation}
where $H[x \vert\,.\,]$ denotes the Fox's $H$-function \cite[Eq.(1.1.1)]{Mathai78}. Substituting \eqref{Tricomi2Meijer} into \eqref{Eq.MGFe2eSNR-2} yields \eqref{Eq.Appendix-C1} shown at the top of the next page. Subsequently, applying the Hankel transform in \eqref{Eq.AppendixA-1} of the product of two Fox's $H$-functions to \eqref{Eq.Appendix-C1} (it is easy to verify that the nine conditions shown in \eqref{Eq.AppendixA-2}-\eqref{Eq.AppendixA-10} are satisfied), we obtain \eqref{Eq.Appendix-C2} at the top of the next page, where $H[^x_y \,\vert\,.\,]$ denotes the generalized Fox's $H$-function of two variables \cite[Eq.(2.1.1)]{Mathai78}. Finally, recalling the relation between the generalized Fox's $H$-function and Meijer's $G$-function of two variables \cite[Eq.(2.3.1)]{Mathai78}, \eqref{Eq.Appendix-C2} reduces to the desired \eqref{Eq.MGFe2eSNR-3}.

\newcounter{mytempeqncnt3}
\begin{figure*}[!t]
\normalsize
\setcounter{mytempeqncnt3}{\value{equation}}
\setcounter{equation}{65}
\begin{eqnarray}  \label{Eq.Appendix-C1}
\mathrm{M}_{\gamma_{e2e}}(s)
= 1-2\sqrt{s}\int_0^\infty{J_1\left(2x\sqrt{s}\right)
H_{1,\,2}^{2,\,1}\left[\frac{x^2}{a_1} \left\vert\begin{gathered} (0, 1) \\(0, 1),\,(1, 1)\end{gathered}\right.\right]
H_{1,\,2}^{2,\,1}\left[\frac{x^2}{a_2} \left\vert\begin{gathered} (0, 1) \\(0, 1),\,(1, 1)\end{gathered}\right.\right]}
\,\mathrm{d}x.
\end{eqnarray}
\setcounter{equation}{\value{mytempeqncnt3}}
\hrulefill
\vspace*{4pt}
\end{figure*}
\setcounter{equation}{66}

\newcounter{mytempeqncnt4}
\begin{figure*}[!t]
\normalsize
\setcounter{mytempeqncnt4}{\value{equation}}
\setcounter{equation}{66}
\begin{equation}  \label{Eq.Appendix-C2}
\mathrm{M}_{\gamma_{e2e}}(s)
= 1-
H_{2,\,[1:\,1],\,0,\,[2:\,2]}^{1,\,1,\,1,\,2,\,2}
\left[\begin{gathered}\frac{1}{a_1\,s}\\ \\ \frac{1}{a_2\,s}\end{gathered}
\left\vert \begin{gathered} (1, 1),\, (0, 1) \\ (0, 1);\,(0, 1) \\ - \\ (0, 1), (1, 1);\,(0, 1), (1, 1)\end{gathered}\right.\right].
\end{equation}
\setcounter{equation}{\value{mytempeqncnt4}}
\hrulefill
\vspace*{4pt}
\end{figure*}
\setcounter{equation}{67}

\section{Proof of Theorem \ref{Theorem1}}
\label{Appendix-D}
Before detailing the proof of our main result in \emph{Theorem~\ref{Theorem1}}, we reproduce one of Gnedenko theorems used later as the following \emph{Lemma~1} \cite[Theorem 2.1.5]{Galambos87}. For more details on various Gnedenko theorems, please refer to \cite[Section 2.1]{Galambos87}.

\begin{lemma}\label{Lemma1}\emph{[Gnedenko's Sufficient and Necessary Conditions for the Domain of Attraction of Weibull Distribution]}
For the parent distribution function $F(x)$ of i.i.d. random sequence $\{X_1, X_2, \cdots, X_n\}$, let its lower endpoint $\alpha(F) \triangleq \inf\{x: F(x) > 0\}$ be finite. Define an auxiliary function $G(x) = F\left(\alpha(F)-\frac{1}{x}\right)$, $x<0$. Assume that there is a constant $\beta > 0$ such that, $\forall x$,
\begin{equation}
\label{Eq.AppendixB-1}
\lim_{t \to -\infty}{\frac{G(tx)}{G(t)}} = x^{-\beta},
\end{equation}
then there are sequences $c_n$, $d_n > 0$ and as $n \to \infty$ the limiting distribution of the minima $W_n \triangleq \min\{X_1, X_2, \cdots, X_n\}$ is given by
\begin{equation}
\label{Eq.AppendixB-2}
\lim_{n \to \infty}\mathrm{Pr}\left\{W_n < c_n+d_{n}x\right\} = L_{2,\,\beta}(x),
\end{equation}
where the Weibull function $L_{2,\,\beta}(x)$ is defined as
\begin{equation}  \label{Eq.AppendixB-3}
L_{2,\,\beta}(x) = \left\{
                     \begin{array}{ll}
                       1-\exp\left(-x^{\beta}\right), & \hbox{if $x > 0$;} \\
                       0, & \hbox{otherwise.}
                     \end{array}
                   \right.
\end{equation}
Also, the normalizing parameters $c_n$ and $d_n$ can be chosen as
\begin{equation}  \label{Eq.AppendixB-4}
c_n = \alpha(F),
\end{equation}
\begin{equation}  \label{Eq.AppendixB-5}
d_n = \sup\left\{x: F(x) \le \frac{1}{n}\right\}-\alpha(F).
\end{equation}
\end{lemma}

Now, we exploit the Gnedenko's sufficient and necessary conditions summarized in the above \emph{Lemma \ref{Lemma1}} to derive the limiting distribution of $F_{\gamma_k}(\gamma)$. At first, it is clear that the lower endpoint of $F_{\gamma_k}(\gamma)$ in \eqref{Eq.CDFgamma-k-IID} is zero, i.e. $\alpha(F) \triangleq \inf\{x: F(x) > 0\} = 0$. In turn, by \eqref{Eq.CDFgamma-k-IID} and performing some algebraic manipulations, the auxiliary function $G(x)$ in \emph{Lemma~\ref{Lemma1}} can be given by
\begin{equation}  \label{Eq.ProofAuxiliaryFunctionG}
G(x) = F_{\gamma_k}\left(-\frac{1}{x}\right) = \frac{1}{1-ax}.
\end{equation}
Then, substituting \eqref{Eq.ProofAuxiliaryFunctionG} into the limit of \eqref{Eq.AppendixB-1} yields
\begin{eqnarray}
\lim_{t \to -\infty}{\frac{G(tx)}{G(t)}}
&  =  & \lim_{t \to -\infty}{\frac{1-at}{1-atx}} \label{Eq.ProofLimit-1} \\
&  =  & \frac{1}{x},                                        \label{Eq.ProofLimit-2}
\end{eqnarray}
where we exploited the L'H\^{o}spital rule to get from \eqref{Eq.ProofLimit-1} to \eqref{Eq.ProofLimit-2}. In view of \eqref{Eq.ProofLimit-2} and \emph{Lemma \ref{Lemma1}}, we obtain that the limiting distribution of $\gamma_{e2e}^\mathrm{upper} = \min_{k=1, \cdots, K}{\gamma_k}$ is the Weibull function $L_{2,\,1}(x)$. Also, according to \eqref{Eq.AppendixB-4} and \eqref{Eq.AppendixB-5}, it is easy to show that the normalizing parameters $c_K = 0$ and $d_K = \frac{a}{K-1} = \frac{\lambda \eta}{(K-1)\sigma^2 }$. Consequently, we have, as $K \to \infty$
\begin{equation} \label{Eq.ProofLimingDistribution}
F_{\gamma_{e2e}^\mathrm{upper}}\left(\frac{\lambda \eta}{(K-1)\sigma^2}\,\gamma\right)
= 1-\exp\left(-\gamma\right),
\end{equation}
which leads to the intended \eqref{Eq.e2eSNRUpperBoundCDF}. Finally, differentiating \eqref{Eq.e2eSNRUpperBoundCDF} with respect to $\gamma$ results in \eqref{Eq.e2eSNRUpperBoundPDF}, which completes the proof.

\vfill


\begin{thebibliography}{99}
\bibitem{FCC2002}
FCC Spectrum Policy Task Force, \emph{Report of the Interference Protection Working Group}, Nov. 15, 2002. Online available at http://transition.fcc.gov/sptf/files/IPWGFinalReport.pdf.

\bibitem{AsghariICC10}
V.~Asghari and S.~A\"{i}ssa, ``Cooperative relay communication performance under spectrum-sharing resource requirements,'' in \emph{Proc. IEEE ICC'10}, pp. 1--6, Cape Town, South Africa, 2010.

\bibitem{MusavianTWC10May}
L.~Musavian, S.~A\"{i}ssa, and S.~Lambotharan, ``Effective capacity for interference and delay constrained cognitive radio relay channels,''  \emph{IEEE Trans. Wireless Commun.}, vol.~9, no.~5, pp.~1698--1707, May~2010.

\bibitem{LeeTWC11Feb}
J.~Lee, H.~Wang, J.~G.~Andrews, and D.~Hong, ``Outage probability of cognitive relay networks with interference constraints,'' \emph{IEEE Trans. Wireless Commun.}, vol.~10, no.~2, pp.~390--395, Feb.~2011.

\bibitem{E1Gamal2011}
A. E1 Gamal and Y.-H. Kim, \emph{Network Information Theory}, Cambridge University Press, 2011.

\bibitem{XiaTCOM12June}
M.~Xia and S.~A\"{i}ssa, ``Cooperative AF relaying in spectrum-sharing systems: performance analysis under average interference power constraints and Nakagami-$m$ fading,'' \emph{IEEE Trans. Commun.}, vol.~60, no.~6, pp.~1523--1533, June 2012.

\bibitem{XiaTCOM12Nov}
M.~Xia and S.~A\"{i}ssa, ``Cooperative AF relaying in spectrum-sharing systems: outage probability analysis under co-channel interferences and relay selection,'' \emph{IEEE Trans. Commun.}, vol.~60, no.~11, pp.~ 3252--3262, Nov. 2012.

\bibitem{HasnaCL03May}
M.~O.~Hasna and M.-S.~Alouini, ``Outage probability of multi-hop transmission over Nakagami fading channels,''  \emph{IEEE Commun. Lett.}, vol.~7, no.~5, pp.~216--218, May~2003.

\bibitem{KaragiannidisTCOM06Jan}
G.~K.~Karagiannidis, T.~A.~Tsiftsis,  and R.~K.~Mallik, ``Bounds for multihop relayed communications in Nakagami-$m$ fading,''  \emph{IEEE Trans. Commun.}, vol.~54, no.~1, pp.~18--22, Jan. 2006.

\bibitem{Kang08VTCspring}
X.~Kang, Y.~C.~Liang, and A.~Nallanathan, ``Optimal power allocation for fading channels in cognitive radio networks: delay-limited capacity and outage capacity,'' in \emph{Proc. IEEE VTC'08-Spring}, pp.~1544--1548,
May 2008.

\bibitem{Etkin07JSAC03}
R.~Etkin, A.~Parekh, and D.~Tse, ``Spectrum sharing for unlicensed bands,'' \emph{IEEE J. Sel. Areas Commun.}, vol. 25, no. 3, pp. 517--528, Apr. 2007.

\bibitem{Lapidoth96TIT09}
A.~Lapidoth, ``Nearest neighbor decoding for additive non-Gaussian noise channels,'' \emph{ IEEE Trans. Inf. Theory}, vol.~42, no.~5, pp.~1520--1529, Sep.~1996.

\bibitem{Musavian08IET06}
L. Musavian and S. Aissa, ``Outage-constrained capacity of spectrum-sharing channels in fading environments,'' \emph{IET Commun.}, vol.~2, no.~6, pp.~724--732, Jul.~2008.

\bibitem{MusavianTWC09Jan}
L.~Musavian and S.~A\"{i}ssa, ``Capacity and power allocation for spectrum-sharing communications in fading channels,'' \emph{IEEE Trans. Wireless Commun.}, vol.~8, no.~1, pp.~148--156, Jan.~2009.

\bibitem{Tse05}
D.~Tse and P. ~Wiswanath, \emph{Fundamentals of Wireless Communications}, Cambridge University Press, 2005.

\bibitem{GoldsmithTIT97Nov}
A.~J.~Goldsmith and P.~P.~Varajya, ``Capacity of fading channels with channel side information,'' \emph{IEEE Trans. Inf. Theory}, vol.~43, no.~6, pp.~1986--1992, Nov.~1997.

\bibitem{ZhaoTSP08Feb}
Q.~Zhao, S.~Geirhofer, L.~Tong, and B.~M.~Sadler,``Opportunistic spectrum access via periodic channel sensing,'' \emph{IEEE Trans. Signal Processing}, vol.~56, no.~2, pp.~785--796, Feb.~2008.

\bibitem{PehaProc09Apr}
J.~M.~Peha, ``Sharing spectrum through spectrum policy reform and cognitive radio,'' \emph{Proc. IEEE}, vol.~97, no.~4, Apr.~2009, pp.~708--719.

\bibitem{Prudnikov86}
A.~P.~Prudnikov, Y.~A.~Brychkov, and O.~I.~Marichev, \emph{Integrals and Series}, Gordon and Breach Science Publishers, 1986.

\bibitem{Gradshteyn07}
I.~S.~Gradshteyn and I.~M.~Ryzhik, \emph{Table of Integrals, Series and Products}, 7th Ed., Academic Press, 2007.

\bibitem{Asghari10}
V.~Asghari, A.~Maaref, and S.~A\"{i}ssa, ``Symbol error probability analysis for multihop relaying over Nakagami fading channels,'' in \emph{Proc. IEEE WCNC'10}, Sydney, Australia, pp.~1--5, Apr.~2010.

\bibitem{XiaTWC11Oct}
M.~Xia, C.~Xing, Y.-C.~Wu, and S.~A\"{i}ssa, ``Exact performance analysis of dual-hop semi-blind AF relaying over arbitrary Nakagami-$m$ fading channels,'' \emph{IEEE Trans. Wireless Commun.}, vol.~10, no.~10, pp.~3449--3459, Oct.~2011.

\bibitem{Mathai78}
A.~M.~Mathai and R.~K.~Saxena, \emph{The $H$-function with Applications in Statistics and Other Disciplines}, Wiley Eastern,~1978.

\bibitem{Ansari11}
I.~S.~Ansari, S.~Al-Ahmadi, F.~Yilmaz, M.-S.~Alouini, and H.~Yanikomeroglu, ``A new formula for the BER of binary modulations with dual-brance selection over gereralized-$K$ composite fading channels,'' \emph{IEEE Trans. Commun}, vol.~59, no.~10, pp.~2654--2658, Oct.~2011.

\bibitem{Hong12TCOMJan}
J.~Hong, B.~Hong, T.~Ban, and W.~Choi, ``On the cooperative diversity gain in underlay cognitive radio systems," \emph{IEEE Trans. Commun.}, vol.~60, no.~1, pp.~209--219,  Jan.~2012.

\bibitem{Galambos87}
J.~Galambos, \emph{The Asymptotic Theory of Extreme Order Statistics}, 2nd Ed., Robert E. Krieger Publishing Co., 1987.

\bibitem{Xia09TVT02}
M.~Xia, Y.~Zhou, J.~Ha, and H.~K.~Chung, ``Opportunistic beamforming communication with throughput analysis using asymptotic approach,'' \emph{IEEE Trans. Veh. Technol.}, vol.~58, no.~5, pp.~2608--2614, June~2009.

\bibitem{Du13TSP}
J.~Du and Y.-C.~Wu, ``Network-wide distributed carrier frequency offsets estimation and compensation via belief propagation,'' \emph{IEEE Trans. Signal Process.}, to appear.

\bibitem{Folk10}
M. Fold, J. H\"{u}sler, and R.-D. Reiss, \emph{Law of Small Numbers: Extremes and Rare Events}, 3rd Ed., Springer Basel, 2010.

\bibitem{WangTCOM03Aug}
Z.~Wang and G.~B.~Giannakis, ``A simple and general parameterization quantifying performance in fading channels,'' \emph{IEEE Trans. Commun.}, vol.~51, no.~8, pp.~1389--1398, Aug. 2003.

\bibitem{Maaref09TCOM01}
A. Maaref and S. A\"{i}ssa, ``Exact error probability analysis of rectangular QAM for single-and multichannel reception in Nakagami-$m$ fading channels,''  \emph{IEEE Trans. Commun.}, vol.~57, no.~1, pp.~214--221, Jan. 2009.

\bibitem{BoradeTIT07Oct}
S.~Borade, L.~Zheng, and R.~Gallager, ``Amplify-and-forward in wireless relay networks: rate, diversity, and network size,'' \emph{IEEE Trans. Inf. Theory}, vol.~53, no.~10, pp.~3302--3318, Oct.~2007.

\end{thebibliography}
\end{document}